\documentclass[namedreferences]{solarphysics}
\usepackage[optionalrh,natbib]{spr-sola-addons} 
\usepackage{graphicx}                    
\usepackage{amssymb}                     
\usepackage{color}                       
\usepackage{url}                         
\usepackage{bm}
\usepackage{lscape}
\usepackage{longtable}

\begin{document}

\begin{article}

\begin{opening}

\title{Coronal Hole Influence on the Observed Structure of Interplanetary CMEs}

%
\author{P.~\surname{M{\"a}kel{\"a}}$^{1,2}$\sep
        N.~\surname{Gopalswamy}$^{2}$\sep
        H.~\surname{Xie}$^{1,2}$\sep
        A.~A.~\surname{Mohamed}$^{1,2}$\sep
        S.~\surname{Akiyama}$^{1,2}$\sep
        S.~\surname{Yashiro}$^{1,2}$
       }

%
\runningauthor{M{\"a}kel{\"a} et al.}
\runningtitle{Coronal Holes and ICME Structure}

%
  \institute{$^{1}$ The Catholic University of America, Washington, DC
 email: \url{pertti.makela@nasa.gov} email: \url{hong.xie@nasa.gov} email: \url{amaal.shahin@nasa.gov}
 email: \url{sachiko.akiyama@nasa.gov} email: \url{seiji.yashiro@nasa.gov}\\
             $^{2}$ NASA Goddard Space Flight Center\\
                     email: \url{nat.gopalswamy@nasa.gov} \\
             }

\begin{abstract}
We report on the coronal hole (CH) influence on the 54 magnetic cloud (MC) and non-MC associated coronal mass ejections (CMEs) selected for studies during the Coordinated Data Analysis Workshops (CDAWs) focusing on the question if all CMEs are flux ropes. All selected CMEs originated from source regions located between longitudes 15E--15W\@. \citet[][\emph{Solar Phys.}, doi:10.1007/s11207-012-0209-0]{xieetal12} found that these MC and non-MC associated CMEs are on average deflected towards and away from the Sun-Earth line respectively. We used a CH influence parameter (CHIP) that depends on the CH area, average magnetic field strength, and distance from the CME source region to describe the influence of all on-disk CHs on the erupting CME\@. We found that for CHIP values larger than 2.6 G the MC and non-MC events separate into two distinct groups where MCs (non-MCs) are deflected towards (away) from the disk center. Division into two groups was also observed when the distance to the nearest CH was less than $3.2 \times 10^{5}$ km. At CHIP values less than 2.6 G or at distances of the nearest CH larger than $3.2 \times 10^{5}$ km the deflection distributions of the MC and non-MCs started to overlap, indicating diminishing CH influence. These results give support to the idea that all CMEs are flux ropes, but those observed to be non-MCs at 1 AU could be deflected away from the Sun-Earth line by nearby CHs, making their flux rope structure unobservable at 1 AU\@.
\end{abstract}

%

\end{opening}

%
\section{Introduction}\label{s:intro}

Coronal mass ejections (CMEs) are magnetized plasma structures that are expelled from the solar corona into interplanetary space. If the CME is launched near the center of the visible solar disk, the CME will hit Earth within few days, possibly causing a severe geomagnetic storm. When the interplanetary counterpart of the CME near the Sun, known as interplanetary CME (ICME), arrives at Earth, an observer near Earth can measure the plasma and magnetic field properties of the passing ICME \citep[see \emph{e.g.},][]{burlagaetal81,gopalswamy06,zurbuchenrichardson06,richardsoncane10}. Therefore if we assume that all CMEs are flux ropes, ICMEs associated with near-disk-center CMEs should show at 1 AU magnetic signatures of flux rope structure, \emph{i.e.}\ smooth rotation of magnetic field. These structures with smoothly rotating magnetic fields are known as magnetic clouds (MCs) \citep[see \emph{e.g.}][]{burlagaetal81,kleinburlaga82}. However observations show that some ICMEs originating from the disk center sources do not have a flux rope structure \citep[see \emph{e.g.},][]{gopalswamy06} and a few even appear to have no ejecta at all \citep{gopalswamyetal09}. A possible solution for this is that the flux rope structure of the ICME exists but cannot be identified from the \emph{in situ} measurements, because the identification of the flux rope signatures becomes more difficult as the spacecraft distance from the flux rope center axis increases \citep[see \emph{e.g.}][]{gopalswamy06,jianetal06,kilpuaetal11}. It is known that the propagation of CMEs is not always radial, indicating that the CME propagation direction must be affected by surrounding coronal structures. The assumption here is that the dominant CME deflection occurs near the Sun, and not later during the ICME propagation in interplanetary space. \citet{wangetal04} have suggested that ICMEs traveling faster than the solar wind speed are deflected to the west and those traveling slower to the east. We do not consider this possible ICME deflection because the solar wind is not fully formed in the height range we are interested in. Already the early white-light observations during the \emph{Skylab} and \emph{Solar Maximum Mission} (SMM) missions revealed that CMEs are deflected towards lower latitudes \citep{hildner77,macqueenetal86}. More recently it has been shown that CME-CME collision \citep{gopalswamyetal01} and CME interaction with coronal holes (CHs) \citep{gopalswamyetal04,gopalwamyetal05} can significantly change the trajectory of the CME\@. Furthermore, \cite{gopalswamyetal09} suggested that CME-CH interaction could explain why no ejecta is observed at 1 AU behind traveling interplanetary shocks that were associated with CMEs launched near the solar disk center, and hence expected to hit Earth. They proposed that combined effects of near-by coronal holes deflect the CME away from the Sun-Earth line, causing the driver behind the shock to miss Earth and the observing spacecraft, resulting in apparently driverless shocks at 1 AU \citep[see also \emph{e.g.},][]{gopalswamy06,jianetal06,rileyetal06}. The CH influence on CMEs was modeled using an \emph{ad-hoc} force depending on the area, average magnetic field strength, and distance of CHs \citep{cremadesetal06}. \cite{mohamedetal12} performed a statistical study that included all disk center CMEs observed by the \emph{Large Angle and Spectrometric Coronagraph} (LASCO) \citep{brueckneretal95} on the \emph{Solar and Heliospheric Observatory} (SOHO) during Solar Cycle 23. They found some evidence supporting the CH influence on the CME propagation. In both studies the measurement position angle (MPA), \emph{i.e.}\ the direction of fastest CME propagation in the sky plane, was used as a proxy of the propagation direction of the CMEs and compared with the position angle (FPA) of the calculated direction of the total CH influence $F$ (see Equation~\ref{eq:F} in Section~\ref{s:ch}).

It this report we study a set of CME-ICME pairs that were especially selected for the two Living With the Star (LWS) Coordinated Data Analysis Workshops (CDAWs) addressing the question if all CMEs are flux ropes or not. The workshops were held in San Diego, USA, in 2010 and in Alcal{\'a} de Henares, Spain, in 2011. We investigate in detail if the geometrical explanation for driverless shocks by \cite{gopalswamyetal09} could also explain why all the selected CDAW ICMEs do not have a flux rope structure at 1 AU even though they originate near the disk center. The idea is that the flux rope structure is not observed because the CME is deflected away from the Sun-Earth line so that the spacecraft at 1 AU crosses the flank of the corresponding ICME \citep[see][]{gopalswamy06}.  We compare the CH influence parameter obtained by \cite{gopalswamyetal09} and \cite{mohamedetal12} with the flux rope fitting results by \citet{xieetal12}. \citet{xieetal12} found that on average the MC associated CMEs are deflected towards and the non-MC associated CMEs away from the Sun-Earth line.

We expect that the CME direction obtained from the flux rope fitting to give a better understanding of the CME propagation direction than the MPA used in the previous studies. The flux rope fitting uses a 3-dimensional model for CMEs and, therefore, results should provide a more realistic estimate of the CME trajectory. Because the MPA is the sky-plane direction of the CME, it cannot describe the 3-dimensional deflection of CMEs accurately, and that can in some cases create problems when the MPA values are compared with the predictions of the CH influence model. For example if a southern polar CH pushes the CME from the southern hemisphere source towards north, but the CME propagation direction still remains in southern hemisphere, the observed MPA value of the CME will be close to 180$^{\circ}$, erroneously indicating a southward deflection of the CME\@.

\section{Data Analysis}\label{s:data}

\subsection{Coronal Mass Ejections}\label{s:cme}
The final data set used in this study consists of 54 CME-ICME pairs of which 23 were labeled as MCs based on the list by R.~Lepping (\url{http://wind.gsfc.nasa.gov/mfi/mag_cloud_pub1.html}). The rest of the events were identified as non-MC events (ejecta). The events were originally selected from a list of CME-driven shocks by \citet{gopalswamyetal10apj}. Selected events were limited to the CMEs that had their solar source location in the longitude range 15E--15W without any limits in the source latitude. The estimated deflection of the CME is based on the analysis by \citet{xieetal12}. They obtained the CME propagation directions by fitting a flux rope model \citep{krallstcyr06} to the white-light images taken by the LASCO experiment on the SOHO\@.

\subsection{Coronal Holes}\label{s:ch}
Coronal holes obtained their name because in the EUV and X-ray images of the solar corona they appear as areas darker than the surrounding corona. However, in images taken at other wavelengths, \emph{e.g.}\ in images taken using the He I 10830 \AA\ line or microwaves \citep{zirker77,gopalswamyetal99}, CHs are brighter than the surrounding solar disk. In the photospheric magnetograms these dark CHs are observed to correspond regions of unipolar magnetic field. It is believed that CHs are filled with open magnetic field lines that extend out into interplanetary space.

In the identification of CH regions and their boundaries we used both EUV images by the \emph{Extreme Ultraviolet Imaging Telescope} (EIT) \citep{delaboudiniereetal95} and photospheric magnetograms by the \emph{Michelson Doppler Imager} (MDI) \citep{scherreretal95}, both instruments on the SOHO spacecraft. First we searched for dark regions in the full-disk EUV 284 \AA\ images and selected for further analysis areas where the EUV intensity was below half of the median EUV intensity of the full solar disk. Filament channels and other interfering dark areas were excluded. Then we looked at the corresponding region in the photospheric magnetogram and defined the CH boundaries to be the boundaries of the major polarity region within the selected region. Further details of the CH identification can be found in \citet{gopalswamyetal09} and \citet{mohamedetal12}.

In the analysis the influence of the CH is described as a force (${\bm f}$) deflecting the CME away from the CH (see Equation~\ref{eq:F}). The direction of this force is assumed to be from the centroid of the CH towards the source region of the CME\@. The magnitude of the force equals the average magnetic field strength ($\langle B \rangle$) within the CH multiplied by the area ($A$) of the CH and divided by the square of the distance ($d$) between the CH centroid and the CME source region. Both the average magnetic field $\langle B \rangle$ and area $A$ of the CH are the line-of-sight corrected values \citep[see][]{gopalswamyetal09,mohamedetal12}. The total force (${\bm F}$) of all CHs on the visible disk is calculated as a vector sum of all CH forces and the magnitude of ${\bm F}$ is called coronal hole influence parameter (CHIP).

\begin{equation}\label{eq:F}
 {\bm F}=\sum_{\rm{CHs}}{{\bm f}}= \frac{\langle B \rangle A}{d^{2}}{\hat{\bm e}},
\end{equation}

where ${\hat {\bm e}}$ is a unit vector pointing from the CH centroid to the CME source region. The corresponding position angle of the $F$ direction is called FPA\@. The unit of the force is Gauss. One should note that this model includes only the possible CH influence. If there are any other mechanisms that deflect CMEs, the model cannot describe their effects or separate the possible CH contribution to the total CME deflection.

\subsection{Data Table}\label{s:datatbl}
We have collected all data used in our analysis into Table~\ref{tbl:1}. The first column of Table~\ref{tbl:1} lists the CDAW event number, the seven next columns give the information about the CME (column 2: date in yyyy/mm/dd format; column 3: time as hh:mm in UT; column 4: source location in heliographic coordinates; column 5: angular distance of the source from the disk center in degrees; column 6: type of the associated ICME; column 7: sky-plane speed in km~s$^{-1}$; column 8: MPA in degrees). Next two columns  list the FPA (column 9) and CHIP (column 10) calculated using our CH influence model. Last two columns are results from the flux rope fitting to the LASCO white-light images of the CME (column 11: propagation direction in heliographic coordinates; column 12: angular distance of the propagation direction from the disk center in degrees). We recalculated the CHIP for the 7 July 2000 (N17) CME because we changed the location of the source region to N04E00. The new CHIP value is 0.3 G\@. We corrected the errors in the calculation of the CHIP values given in \citet{mohamedetal12} for 4 events: N08 (CHIP=1.1 G), N24 (CHIP=6.0 G), N28 (CHIP=12.0 G), and N32 (CHIP=5.7 G).

\section{Results}

\subsection{CME Deflection by Coronal Holes}\label{s:deflection}

\begin{figure}
 \centerline{\includegraphics[width=0.9\textwidth,clip=]{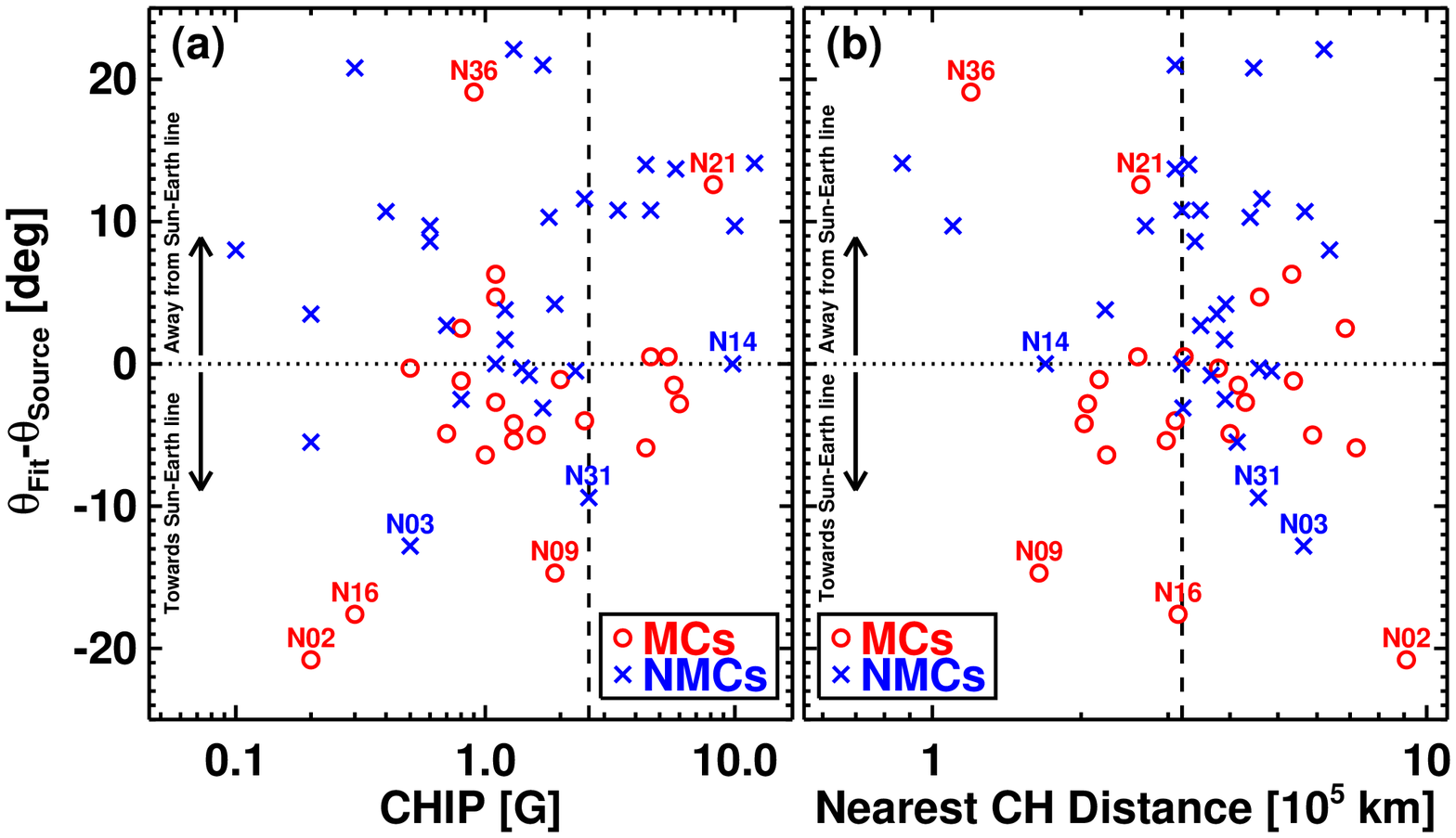}}
 \caption{Deflection of the CME direction relative to the Sun-Earth line based on the flux rope fitting by \citet{xieetal12} as a function of (a) the CHIP (b) the distance of the nearest CH\@. $\theta_{\rm{Source}}$ and $\theta_{\rm{Fit}}$ are angular distance of the source location and the CME propagation direction from flux rope fitting relative to the disk center, respectively. Blue crosses (rec circles) mark MC (non-MC) events. Dashed lines in Figures (a) and (b) mark the CHIP value of 2.6 G and the distance of $3.2 \times 10^{5}$ km respectively. The event number is plotted next to few selected data points. These events are discussed in more detail in the text.}\label{fig:CHIPDist}
 \end{figure}

In order to characterize the CME deflection from the radial propagation we calculated the angular distances from the Sun-Earth line, \emph{i.e.}\ from the disk center, for the CME source region ($\theta_{Source}$) and the flux rope propagation direction ($\theta_{Fit}$). The CME source regions and flux rope propagation directions were taken from Table~1 by \citet{xieetal12}. \citet{xieetal12} used white-light images by the SOHO/LASCO coronagraph to forward model the flux rope orientation and propagation near the Sun. In Figure~\ref{fig:CHIPDist} we have plotted the estimated CME deflection $\theta_{Fit} - \theta_{Source}$ {\it vs.}\ the CHIP (Figure~\ref{fig:CHIPDist}a) and the nearest CH distance (Figure~\ref{fig:CHIPDist}b). The negative values of $\theta_{Fit} - \theta_{Source}$ indicate deflection towards the Sun-Earth line, \emph{i.e.}\ it is more likely that an observer at 1 AU should detect an MC structure, assuming that all CMEs are flux ropes. The red circles (blues crosses) mark MC (non-MC) events respectively.

The CHIP values plotted in Figure~\ref{fig:CHIPDist} are from Tables~2 and 3 in the paper by \citet{mohamedetal12} that includes data from \citet{gopalswamyetal09}. Some data values were recalculated as mentioned in Section~\ref{s:datatbl}. The nearest CH distances are extracted from the data used in the calculations of CHIP values. In the plots and our discussions we have excluded the event N13 on 1999 September 20 because it is a very faint halo CME with an exceptionally large CHIP value due to a large CH at SW from the source region at S20W05. As discussed by \citet{mohamedetal12} this large CH should push the CME towards the NE direction. The measurement position angle (MPA) for this event is 14$^{\circ}$, indicating that the fastest part of the CME travels approximately to the NE direction as expected. However the result from the flux rope fitting for this event shows no deflection at all, \emph{i.e.}\ the CME should appear to propagate radially towards south. We think that the faintness of the CME makes this event very difficult to accurately fit with a flux rope model, therefore we have excluded it from our analysis.

The general conclusion from Figure~\ref{fig:CHIPDist} is that the majority of MCs (red circles) lie below the dotted horizontal line marking the zero deflection. This indicates that MCs are favorably deflected towards the Sun-Earth line as discussed in the paper by \citet{xieetal12}. Only 7 out of 23 (30\%) MC events appear to have  been deflected further away from the Sun-Earth line, and only 2 out of the 7 events (events N21 and N36) are deflected more than 10$^{\circ}$ away from the Sun-Earth direction. Similarly most of the non-MC events (blue crosses) lie above the zero level, \emph{i.e.}\ they are deflected away from the disk center. Only 8 non-MC events out of 30 (27\%) are deflected towards the disk center. The non-MC events N03 and N31 are the two extreme events out of the 8 events with a towards-disk-center deflection. There are two non-MC events with no deflection.

When considering the CH influence and how well our model of that can explain the estimated CME deflection we notice in Figure~\ref{fig:CHIPDist}a that the range of CHIP values can be divided roughly into two regions. Events with CHIP larger than 2.6 G show clear separation of MC and non-MC deflection. The MC events are deflected towards the Sun-Earth line and non-MC event away, with two exceptions: the non-MC event N14 with CHIP=9.8 G and no observed CME deflection and the MC event N21 with CHIP=8.2 and deflection away from the Sun-Earth line. We will discuss these two events below in detail. In the CHIP value region less than 2.6 G the deflection distributions of the MC and non-MC events are overlapping. Most of the events are concentrated between $-10^{\circ} < \theta_{Fit} - \theta_{Source} < 10^{\circ}$. But still in the region of lower CHIP values the majority of MC events (13 out of 17 or 76\%) are deflected towards disk center and that of non-MC events (14 out of 23 or 61\%) are deflected away from the disk center. The 3 more extreme cases are MC events N02 and N16 that show larger deflection even though both events have the low CHIP value, and the event N09 with a moderate CHIP value of 1.9 G\@. We discuss also these events later in this section.

In Figure~\ref{fig:CHIPDist}b we have plotted the nearest CH distance {\it vs.}\ the CME deflection. Also the range of the nearest CH distances can be divided into two regions with differing CME deflection distributions. Again the events N21 and N36 form an exception discussed later. If the distance to the nearest CH is less than $3.2 \times 10^{5}$ km the groups of the MC and non-MC events are clearly separated. This distance corresponds to approximately a quarter of the solar radius. When the CH distance increases the deflection distributions start to overlap. We did not find similar clear division for the area times the average magnetic field strength of CHs. Clearly the distance of the nearest CH is a significant factor for the CH influence.

Figure~\ref{fig:CHIPDist} also shows that the magnitude of CME deflection for MC events is confined between -7$^{\circ}$$\leq \theta \leq$7$^{\circ}$, only 5 MC events (event number plotted next to the data point in Figure~\ref{fig:CHIPDist}) show larger values of CME deflection. On the other hand, non-MC events have CME deflection values that are scattered into a wider range between -6$^{\circ}$$\leq \theta \leq$23$^{\circ}$. In the region where CHIP value are larger than 2.6 G the non-MC events have CME deflection values near and above 10$^{\circ}$, which supports the idea that CHs influence the CME propagation. The CME deflection values for the MC events do not show similar shift at large CHIP values. On the other hand the number of MC events in this region is low, so this could be by chance. It is also possible that it is an intrinsic characteristic of the CME population associated with MCs to be less deflected than those associated with the non-MCs.

\subsection{18 October 1999 Non-MC Event}\label{s:nmc14}
\begin{figure}
\begin{center}
\includegraphics[width=0.3\linewidth,clip=]{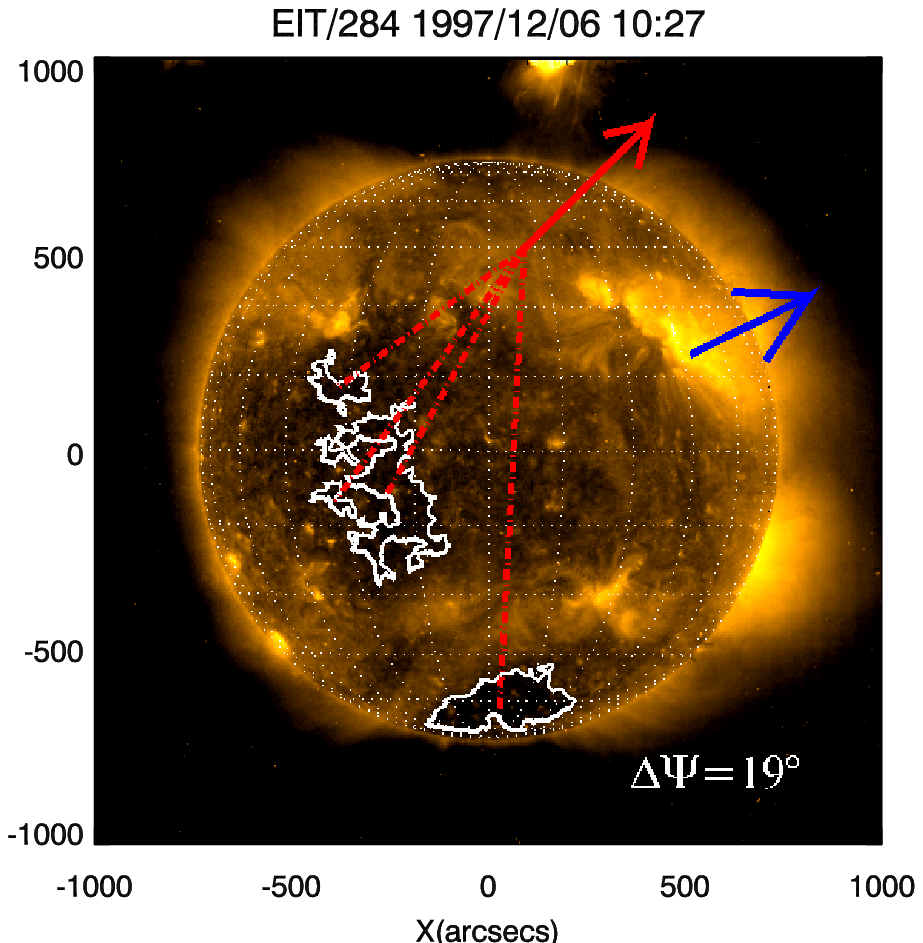}
\includegraphics[width=0.3\linewidth,clip=]{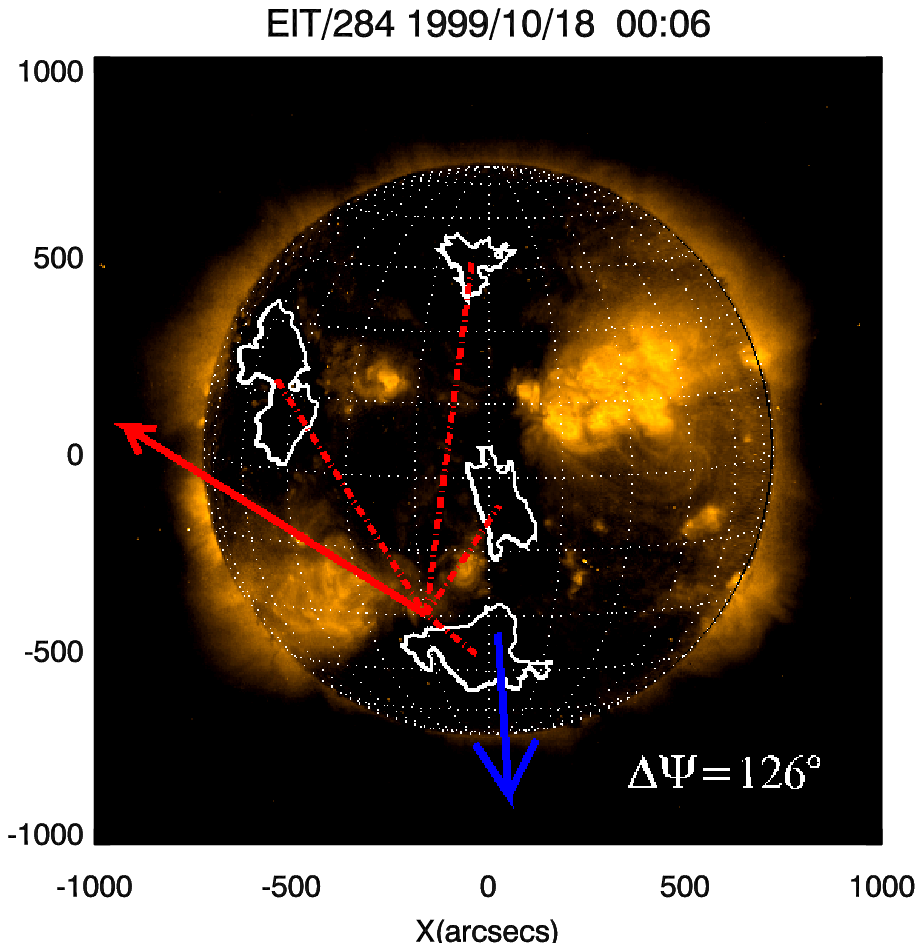}
\includegraphics[width=0.3\linewidth,clip=]{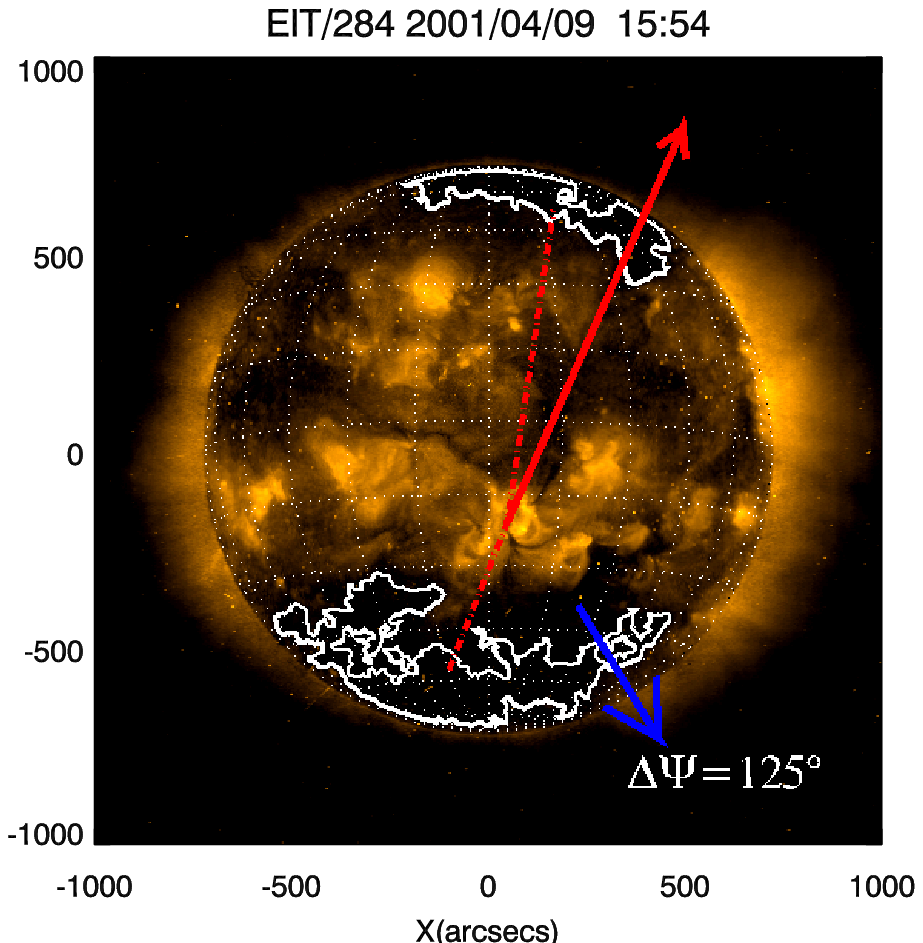}
\end{center}
\caption{EUV images showing the coronal holes for the 6 December 1997 (left), 18 October 1999 (middle), and 9 April 2001 (right) non-MC events discussed in the text. Blue arrow marks the MPA of the CME, red arrow the direction of the CH influence and $\Delta \Psi$ is the angle between them. Figures from \citep{mohamed11}.}\label{fig:NMC}
\end{figure}

The CME (N14) in question is a faint and slow partial halo on 18 October 1999 at 00:06 UT with MPA=184$^{\circ}$ at S30E15. Another narrow and slow CME with MPA=40$^{\circ}$ occurs at the same time as the partial-halo CME\@. The conclusion that these events are two separate CMEs is based on a slight difference in speeds of the emerging loop structures towards the NE and S directions, but the sequence of events is complex and open to interpretations. This event is difficult to fit with a flux rope model accurately. The locations of the selected CHs are shown in Figure~\ref{fig:NMC} (middle). Therefore we consider the CME deflection in this event to be uncertain.

\subsection{6 December 1997 Non-MC Event}\label{s:nmc03}
The CME (N03) on 6 December 1997 at 10:27 UT has the largest deflection towards the disk center of all non-MC events, but the estimated CHIP value is only 0.5 G\@. On the other hand the CME source region is farthest away from the Sun-Earth line of all events. The estimation of the source location for this event is somewhat complicated. The first indication of the possible CME eruption was an eruptive prominence at high northern latitude (N45W10), followed by a formation of large arcades about four hours later near active regions 8115 and 8113. We selected the source to be at N45W10, but \citet{xieetal12} used in their calculations the later source location at N25W40. In either case the angular distance of the source for this event is over 40$^{\circ}$ from the disk center, making it the most distant event of all the CDAW events. The fitted propagation direction of the CME is N15W30, which is still about 33$^{\circ }$ from the disk center. Therefore, even though the CME was deflected towards the Sun-Earth line, the distance from the Sun-Earth line remained large. The MPA of the CME is 315$^{\circ}$, indicating propagation in the NW quadrant. If the location used in flux rope fitting is accurate then the event is beyond the longitudinal range of our study. Because this latter location is so close to the western limb the calculations are unreliable because they do not include the influence from the possible CHs near the western limb (see Figure~\ref{fig:NMC} left). The Kitt Peak CH map shows near the west limb a long, narrow elongation of the northern polar CH reaching almost to the N20 latitude. In the EUV images this CH appears to be masked by the bright loops in the forefront. Because of these ambiguities in the location of the source region and CHs we must consider this event uncertain.

\subsection{9 April 2001 Non-MC Event}\label{s:nmc31}

The halo CME (N31) on 9 April 2001 at 15:54 UT is another non-MC event with $\sim 10^{\circ}$ deflection towards the Sun-Earth line. This event is the one which we used as a limit between the two CHIP ranges, so the corresponding CHIP value is 2.6 G\@. The CME occurred at S21W04 and the fitted propagation direction is S12E01. The MPA angle of the CME is 221$^{\circ}$, again indicating that the CME propagated towards south. The nearest CH was the southern polar CH, which should push the CME towards north exactly as the flux rope fitting indicated (Figure~\ref{fig:NMC} right). The question then is why this event was classified as a non-MC event even though its propagation direction was less than 10$^{\circ}$ from the Sun-Earth line? It is quite possible that the associated ICME was misidentified because the CME was followed by another faster halo CME (N32) from the same region on 10 April 2001 at 05:30 UT\@. The speed of the 9 April CME was 1192 km~s$^{-1}$ and the 10 April CME had a speed twice of that of the preceding CME, 2411 km~s$^{-1}$. The corresponding shocks were detected only two hours apart at 14:12 UT and 16:19 UT respectively. Therefore, we think that the corresponding ICMEs were merging and the flux rope structure of the 9 April CME was destroyed in the process. This means also that our CHIP limit could be lowered to $\sim$2 G\@.

\subsection{12 May 1997 MC Event}\label{s:mc02}
\begin{figure}
\begin{center}
\includegraphics[width=0.3\linewidth,clip=]{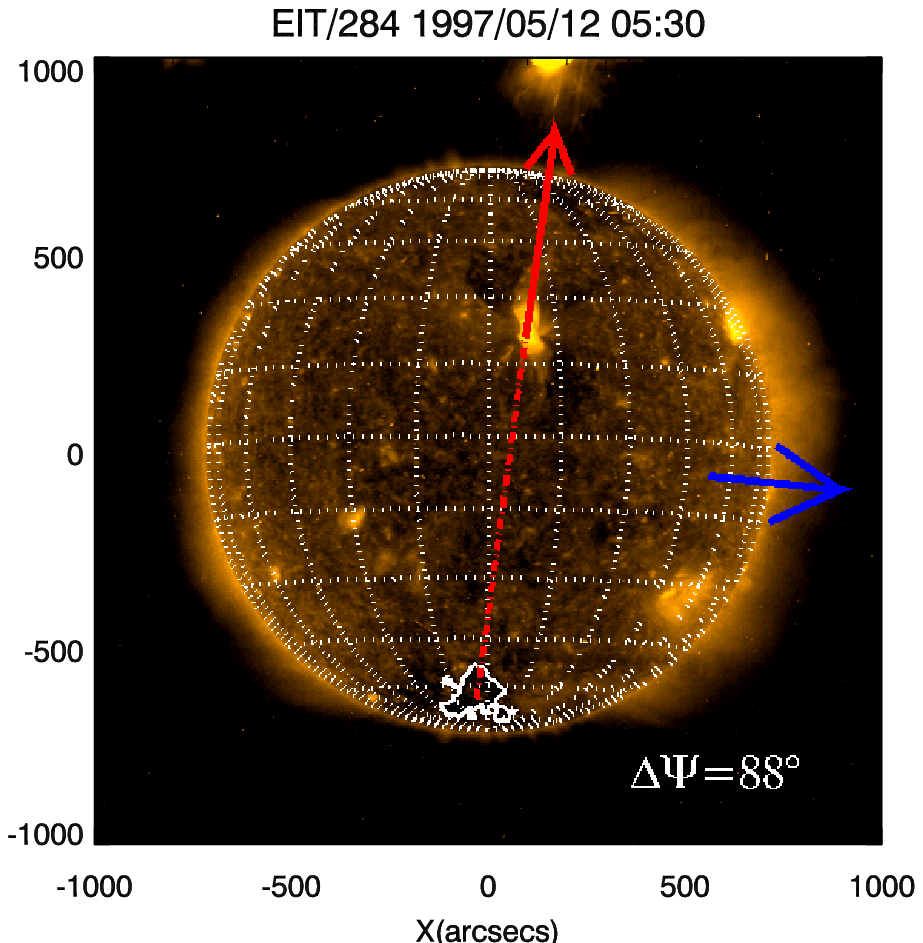}
\includegraphics[width=0.3\linewidth,clip=]{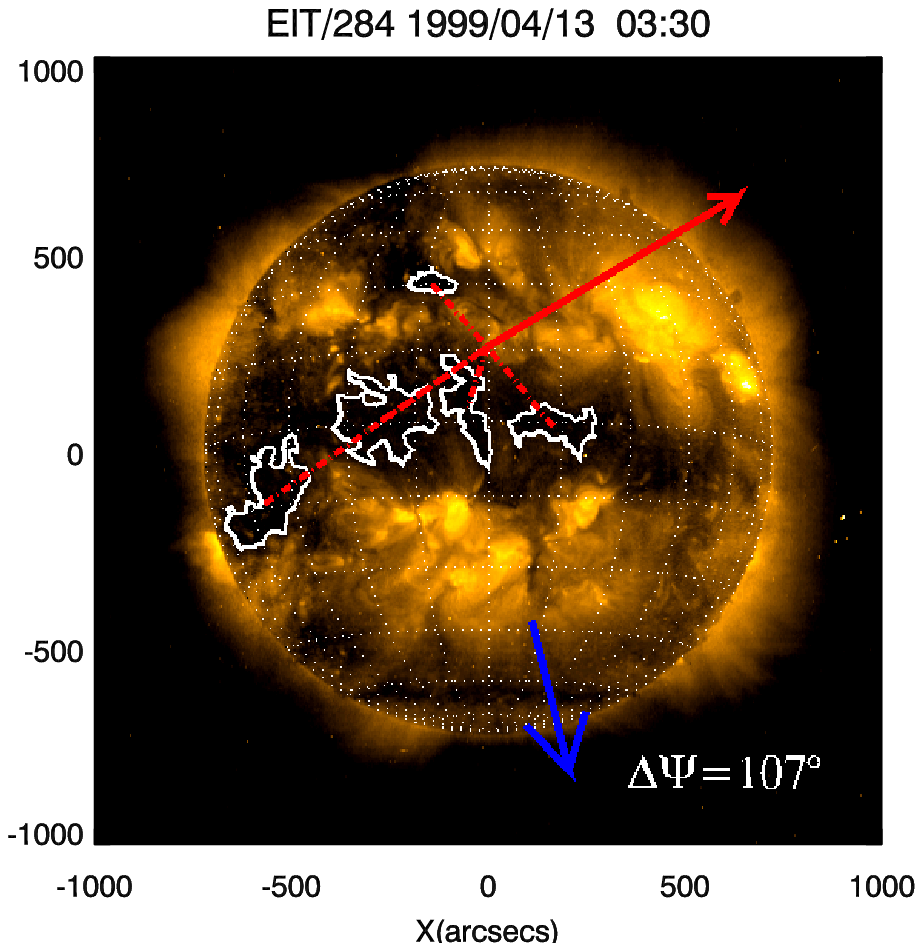}
\includegraphics[width=0.3\linewidth,clip=]{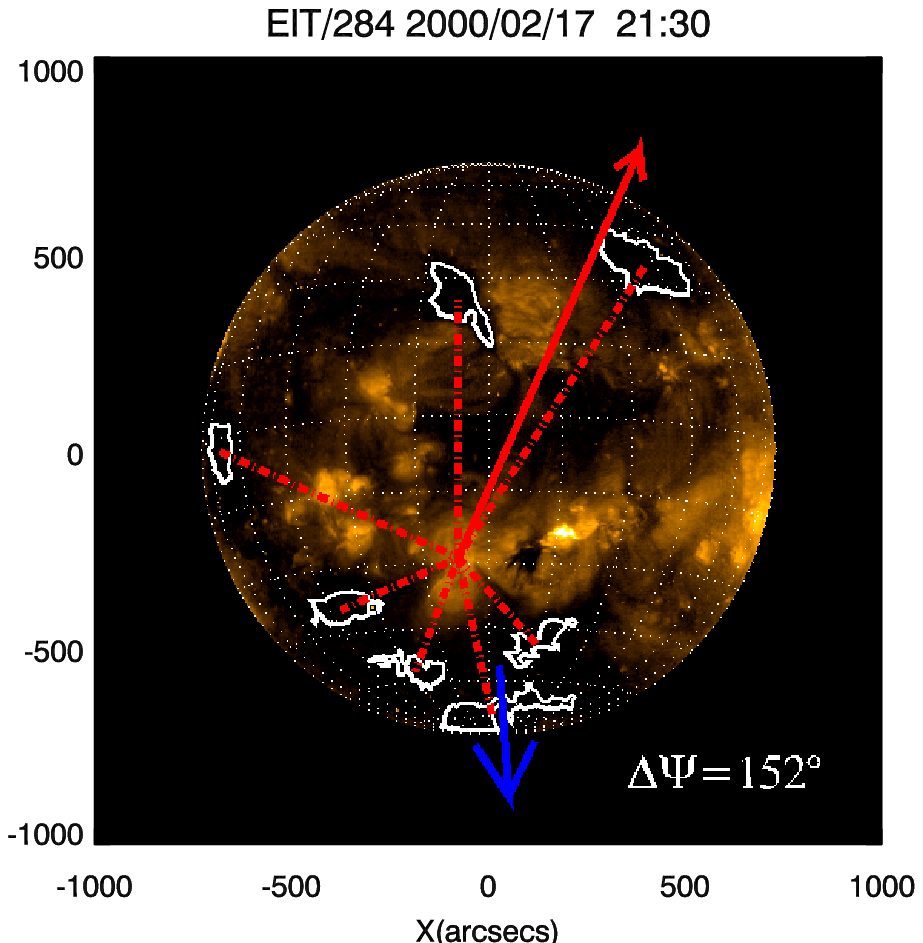}\\
\includegraphics[width=0.3\linewidth,clip=]{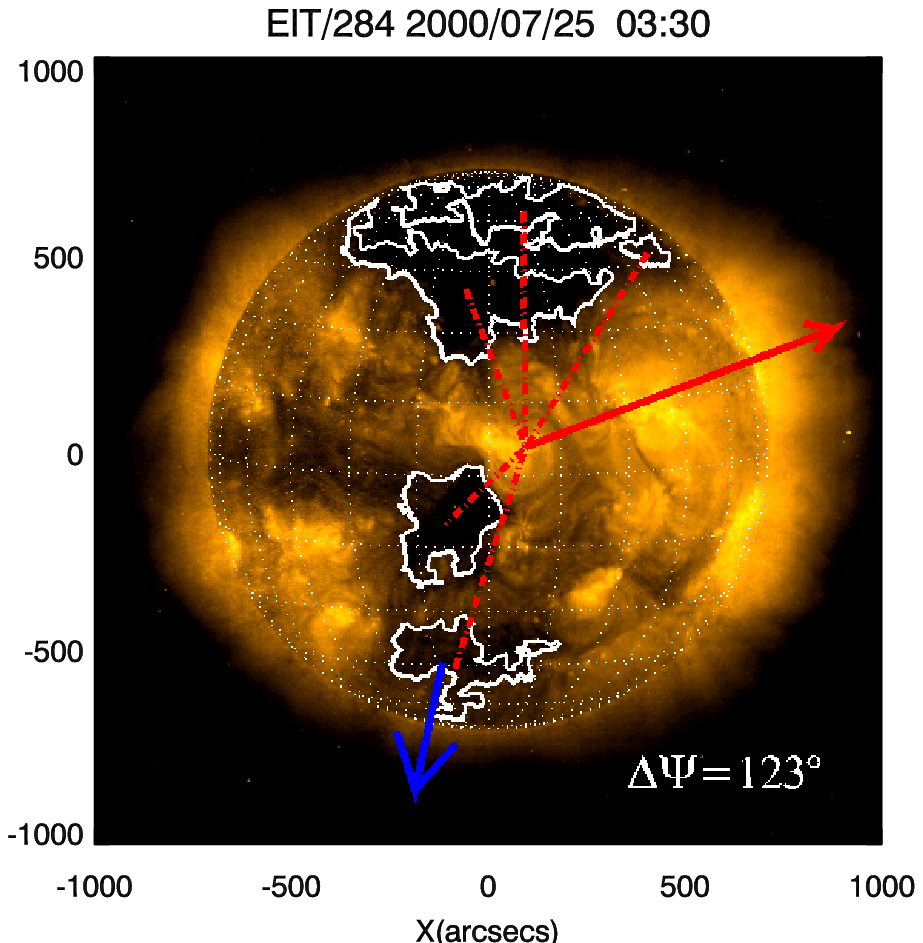}
\includegraphics[width=0.3\linewidth,clip=]{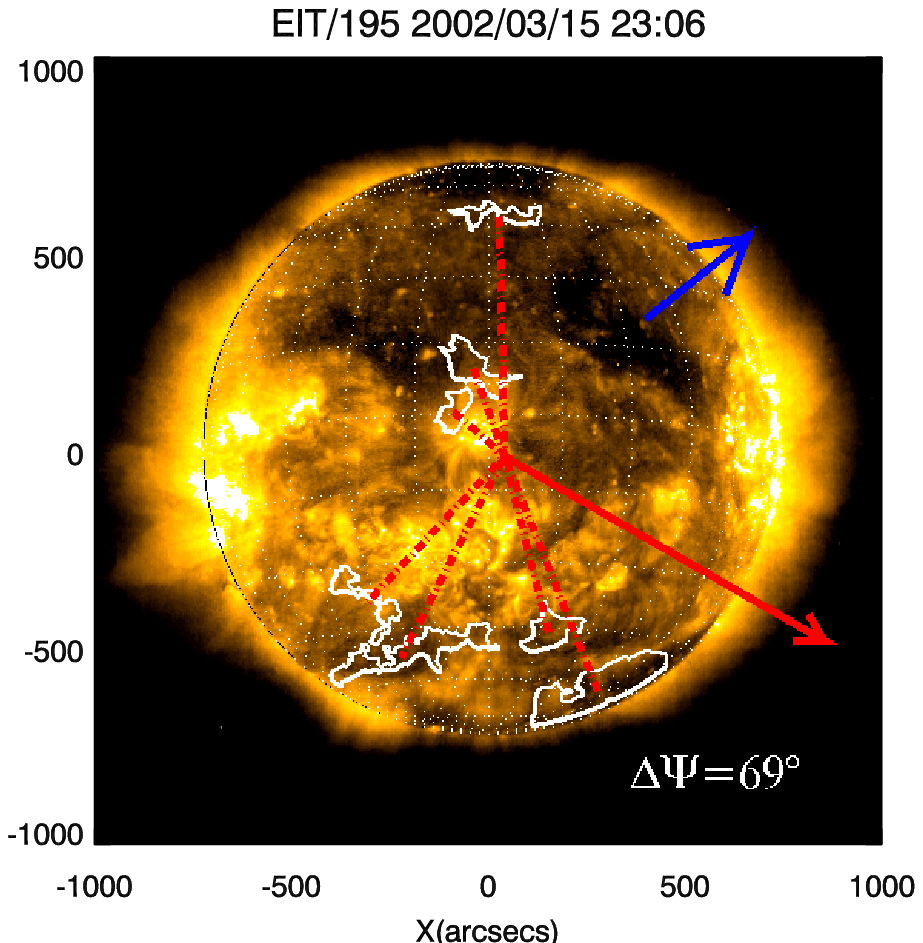}
\end{center}
\caption{EUV images showing the coronal holes for the 12 May 1997 (top left), 13 April 1999 (top middle), 17 February 2000 (top right), 25 July 2000 (bottom left), and 15 March 2002 (bottom left) MC events discussed in the text. Blue arrow marks the MPA of the CME, red arrow the direction of the CH influence and $\Delta \Psi$ is the angle between them. Figures from \citep{mohamed11}.}\label{fig:MC}
\end{figure}

The slow halo CME (N02) on 12 May 1997 at 05:30 UT occurred at N21W08 during the period of minimum solar activity. We identified only one polar CH far away in the southern pole as is typical during a solar minimum (Figure~\ref{fig:MC} top left). Therefore the CHIP value we obtained is only 0.2 G\@. We do not expect the southern polar CH have any significant influence on the CME propagation. The CME propagation direction was N01W02, only 4$^{\circ}$ from the disk center. So what explains the relatively large deflection of the CME towards the disk center? The possible cause is the global solar magnetic field, which during the solar minimum is a well-organized dipole field associated with strong magnetic fields in the polar CHs that are known to exist in the polar regions during solar minimum. CME deflection towards lower latitudes during solar minima was first observed by \citet{hildner77} and \citet{macqueenetal86}. Probably our CH selection method cannot identify the near-limb northern polar CH if it has only a small area on the visible side of the Sun. Therefore the large CME deflection in this case can be attributed to the effects of the large-scale solar magnetic field configuration due to polar CHs \citep[see also \emph{e.g.},][]{gopalswamythompson00,filippovetal01,plunkettetal01,cremadesetal06,kilpuaetal09}.

\subsection{13 April 1999 MC Event}\label{s:mc09}
The partial-halo CME (N09) on 13 April 1999 occurred at 03:30 UT when the solar X-ray emission was low. The CME was relatively faint with an uncertain width ($>261^{\circ}$), and it was expanding fastest towards south (MPA=194$^{\circ}$). The selected source for this event is a disappearing filament at N16E00 with the post-flare arcade loops reaching a B3.4 class in the X-ray intensity. However, another possible source candidate for this event is an EUV dimming at S13E21. The fitted flux rope direction was S02W06 matching with the observed MPA towards south. The three nearest CHs were located in the SE, S, and SW direction from the selected source (see Figure~\ref{fig:MC} top middle), and they were of average size. The calculated CHIP value for this event is 1.9 G with the direction towards NW\@. The large deflection obtained by the flux rope fitting suggests that there was a CH near the source as observed, but the calculated direction of the CHIP deviates significantly from the expected deflection towards south. A better agreement with observations would be achieved if the source was located south of the CHs as suggested by the EUV dimming. But if the source was located at S13E12 then the CME propagation direction from the flux rope fitting (S02W06) would indicate deflection towards north. Because we identified a total of 4 low-latitude CHs, which is an unusually large number, it is possible that our CH selections in this case are not correct. The inspection of the EUV image shows bands of bright regions at mid-latitudes in the southern and northern hemisphere and a dark equatorial region in between. A few long and faint transequatorial loops appear to connect these bright regions in south and north, so it could be that the magnetic field lines are closed in the equatorial region. Otherwise unless we have misidentified the associated CME, which seems unlikely, this event is difficult to explain based on the CH influence model.

\subsection{17 February 2000 MC Event}\label{s:mc16}
The halo CME (N16) on 17 February 2000 at 21:30 UT was launched from a location at S29E07. The fitted propagation direction was S12W02 and the MPA was 184$^{\circ}$. The estimated CHIP values was 0.3 G\@. We identified 3 relatively small CHs at almost equal distance (3.1--3.8$\times 10^{5}$ km) at SE, S, and SW directions from the source (Figure~\ref{fig:MC} top right). In addition there was a southern polar CH (5.5$\times 10^{5}$ km) and another 3 CHs were in NE, N, and NW directions but at large distances (6.3--8.8$\times 10^{5}$ km). The calculated force ${\bm F}$ was towards N-NE direction, which coincides well with the result from flux rope fitting. Why then our CHIP value is so low if the CME was deflected by the nearby CHs? Because CHs are surrounding the CME source region in all directions, the low CHIP value might be an indication that our simple model overestimates the influence of CHs far away from the source region. Considering that Sun is a sphere it is possible that the influence of CHs more than solar radius away is less than predicted by our model.

\subsection{25 July 2000 MC Event}\label{s:mc21}
The 25 July 2000 CME (N21) at 03:30 UT occurred at N06W08, but the flux rope was observed first above the southern solar limb (MPA=168$^{\circ}$). Also the flux rope fitting shows deflection towards the S-SE direction, as the fitted propagation direction was S15E04. However, from the Figure~\ref{fig:MC} (bottom left) we see that the nearest CH on disk was in the SE direction from the source region. This CH had also a strong average magnetic field ($\langle B \rangle =$11.9 G). Our calculations show that this CH dominates the total influence of all the on-disk CHs and therefore the CME should be pushed towards the NW direction. There were large northern polar CHs relatively nearby ($\langle B \rangle =$2.6--4.8 G) but according to our estimation their effect did suffice only to turn the total CH influence direction slightly towards west. However, this CME has an uncertain source region. There was a M8.0 flare eruption at 02:43 UT in the active region 9097 at N06W08 and approximately at the same time an eruptive prominence occurred further south at S14W04. Interestingly the flux rope fitted propagation direction of the CME is close to this eruptive prominence location. Therefore, we conclude that the complex events at the Sun make the identification of the CME source location uncertain, and according to our CH influence model the eruptive prominence would be a more favored source of the CME\@.

\subsection{15 March 2002 MC Event}\label{s:mc36}
The 15 March 2002 event (N36) was a halo CME first observed at 23:06 UT expanding towards the NW direction (MPA=309$^{\circ}$). Its solar source was the M2.2 flare at S08W03 at 22:09 UT\@. Our CH identification found multiple small CHs scattered around the solar disk (see Figure~\ref{fig:MC} bottom right). The influence of each CH was calculated to be relatively weak ranging from 0.2 G to 1.6 G, resulting in a weak total influence (CHIP=0.9 G) pushing the CME towards the SW direction. The fitted propagation direction was N15W01, so the CME appears to be deflected towards north. In this case there is no ambiguity in the source location. In addition, the solar south pole is inclined towards Earth in mid March, so the CME source is very close to the disk center. Therefore, we cannot expect the deflection away from the Sun-Earth line will necessarily mean that observer at 1 AU cannot detect the flux rope structure of the CME\@. Question remains what caused this deflection. There could be some uncertainty in the identification of the CHs, because there is a nearby filament channel north of the source region extending from the central meridian towards the E-NE direction. This filament channel might interfere with the identification of the two CHs near the CME source region. In any case, it appears that there is no clear CH close enough in the S-SE direction from the source region, which could push the CME towards the N-NW direction. However, this event occurred during the solar maximum, when the solar magnetic field is very complex, so distorted local magnetic structures could result in the CME deflection.

\subsection{FPA versus the CME Propagation Direction}\label{s:fpa}

\begin{figure}
\centerline{\includegraphics[width=1.0\textwidth,clip=]{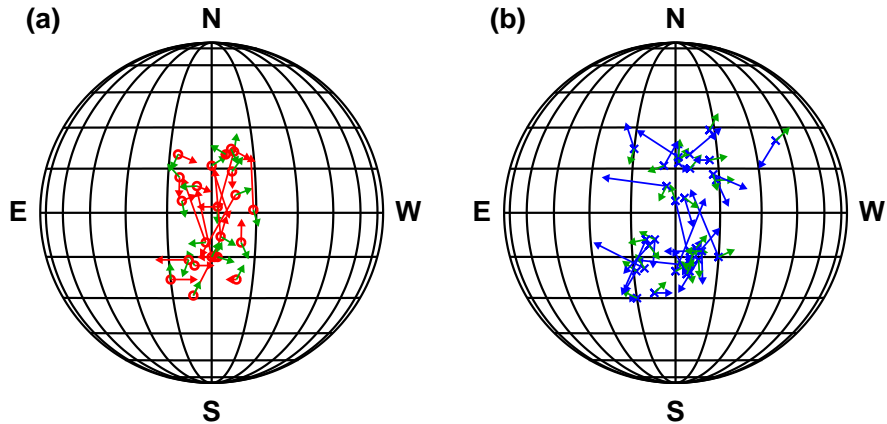}}
\caption{FPA from the CH deflection model (Table~\ref{tbl:1}, column 9) compared with the shift of the CME propagation direction by \citet{xieetal12} relative to the source location. The red (a) and blue (b) arrows correspond to MC and non-MC events. The red and blue arrows start at the CME source location (Table~\ref{tbl:1}, column 4) and end at the CME propagation direction (Table~\ref{tbl:1}, column 11). The green arrows show the direction of the CHIP\@.}\label{fig:FPA}
\end{figure}

In Figure~\ref{fig:FPA} we have plotted the FPA angles together with arrows that start from the source region and end at the estimated CME propagation direction from the flux rope fitting by \citet{xieetal12}. As can be seen the alignment of the green arrows (FPA) with the red (MCs) and blue (non-MCs) is not particularly good. The CH influence model clearly provides less accurate estimates for the FPA than for the CHIP\@. We believe that the moderate correspondence of the FPA with the shift of the CME propagation direction relative to the source location is partly due to simplified description of the CH as a single point (the centroid of the CH area) in the model. Especially when the CH has very elongated shape, the CH section nearest to the source most likely contributes more in the CME deflection than the rest of the CHs. Figure~\ref{fig:CHIPDist}b showed that the distance of the CH is significant parameter for the CME deflection. In addition the identification of CH areas has uncertainties. As discussed by \citet{gopalswamyetal09} the selection method of the CH area may not fully select open field regions due to the foreground coronal emission. Other features on the Sun can interfere with the CH identification as might be the case in the 13 April 1999 event we discussed in Section~\ref{s:mc09}, where long faint transequatorial loops possibly interfered with the CH identification resulting in unusually large number of CHs.

In our study we have not considered the possible uncertainties in the fitting of the flux rope model. When the CME appears very faint in the coronagraphic images or when parts of the successively launched CMEs overlap each other, the identification of features of the CME becomes difficult. This results in not easily quantifiable uncertainties in the fitting of the flux rope model and in the estimation of the CME propagation direction.

\section{Summary and Conclusions}\label{s:summary}

Our report is one of the contributions originating from collaborations during the LWS CDAW meetings focusing on the question if all CMEs are flux ropes. In a recent study \citet{gopalswamyetal09} analyzed ICMEs that originated from disk-center sources and therefore were expected to be directed towards Earth, but which did not have an observed ejecta at 1 AU\@. As an explanation they suggested that nearby CHs pushed the CMEs near the Sun away from the Sun-Earth line far enough that the driving ejecta of the corresponding ICME became unobservable at 1 AU\@. \citet{gopalswamyetal09} described the CH influence on CMEs by using a simple parameter called CHIP (see Equation~\ref{eq:F}) depending on the area, average magnetic field and distance from the source of the CH \citep[see also][]{cremadesetal06}. \citet{mohamedetal12} performed an expanded statistical study of the CH influence on CMEs during the whole Cycle 23. They found some evidence supporting the CH influence. In our study we have utilized results from the flux rope fitting reported by \citet{xieetal12}, who found that on average CMEs associated with MCs are deflected closer to the disk center and those associated with non-MCs away from the disk center.

When we compared the CME deflection \citep{xieetal12} to the CH influence parameter CHIP \citep{gopalswamyetal09,mohamedetal12} we found support to the CH influence as described by CHIP on the CME propagation. We found that for the CHIP values larger than 2.6 G the CME deflection distributions are divided into two separate groups where the MCs are deflected towards and non-MCs away from the Sun-Earth line. At CHIP values lower than 2.6 G the deflection distributions of MC and non-MC events overlap but still the average deflection direction for MC is towards and non-MCs away from the Sun-Earth line. We also found that the deflection as a function of the distance of the nearest CH is divided into two distance regions. If the nearest CH is closer than $3.2 \times 10^{5}$ km from the CME source region, the deflection distributions of the MCs and non-MCs again are separated into two groups: MCs are deflected towards the Sun-Earth line and non-MCs correspondingly away. When the distance to the nearest CHs increases the CH influence on CMEs decreases and the deflection distributions start to overlap. This indicates that the distance to the nearest CH is an important parameter for the CME deflection. We also found the scatter of the CME deflection values to be larger for non-MC event than for the MC events.

There were few events that had exceptionally large values of the CME deflection, which we discussed in more detail. Most of the events revealed unavoidable uncertainties in identifying CME solar sources and CHs using the methods applied here, and which resulted in uncertain predictions of the CH influence. In addition to the problems in the identification of the features on the Sun, the model used to calculate the CHIP reduces the CH to a single point (the centroid of the CH). This assumption is incorrect especially if the CH has a very elongated shape, because then the nearest section of the CH to the CME source is the most likely area pushing the CME. During the solar minimum the global dipole magnetic field due to strong magnetic fields in the polar CHs deflects CMEs towards the lower latitudes \citep[see \emph{e.g.},][]{hildner77,macqueenetal86,gopalswamythompson00,filippovetal01,plunkettetal01,cremadesetal06,kilpuaetal09}. During solar maximum the solar magnetic field configuration can be very complex, so that local magnetic structures near the CME source may direct the CME to propagate non-radially. We cannot separate or exclude these other effects in our calculations.

As a final point we like to mention that the CHIP estimates might improve if one modifies Equation~\ref{eq:F} so that the CH force is proportional to the square of the average magnetic field strength of the CH as suggested by \citet{gopalswamyetal10aip}. They proposed this modification because $B^{2}$ represents magnetic pressure and therefore could be a better CH parameter in the calculations of the CH influence.

In summary, we found evidence by using a simple CH influence model that CHs probably deflect CMEs and that the deflection pattern of the MC and non-MC associated CMEs near the Sun as reported by \citet{xieetal12} is at least partly explained by the CH influence.

%
\begin{landscape}
\begin{longtable}{lccccrcrrcrrccr}
\caption{List of the CDAW events and their CH influence and best-fit flux rope parameters.}\\

\hline N & & \multicolumn{7}{c}{CME} & & \multicolumn{2}{c}{CHIP$^{a}$} & & \multicolumn{2}{c}{Flux Rope Fitting$^{b}$}\\
\cline{3-9} \cline{11-12} \cline{14-15}
 & & Date & Time & Source & \multicolumn{1}{c}{$\theta_{Source}$} & Type & \multicolumn{1}{c}{Speed} & \multicolumn{1}{c}{MPA} & & \multicolumn{1}{c}{FPA} & \multicolumn{1}{c}{F} &  & Direction & \multicolumn{1}{c}{$\theta_{Fit}$}\\
 & & & [UT] & & \multicolumn{1}{c}{[deg]} & & \multicolumn{1}{c}{km s$^{-1}$} & \multicolumn{1}{c}{[deg]} & & \multicolumn{1}{c}{[deg]} & \multicolumn{1}{c}{[G]} & & & \multicolumn{1}{c}{[deg]} \\ \hline
\endfirsthead

\multicolumn{15}{l}{\small\sl \tablename\ \thetable{} -- continued from previous page}\\ \hline
N & & \multicolumn{7}{c}{CME} & & \multicolumn{2}{c}{CHIP} & & \multicolumn{2}{c}{Flux Rope Fitting}\\
\cline{3-9} \cline{11-12} \cline{14-15}
 & & Date & Time & Source & \multicolumn{1}{c}{$\theta_{Source}$} & Type & \multicolumn{1}{c}{Speed} & \multicolumn{1}{c}{MPA} & & \multicolumn{1}{c}{FPA} & \multicolumn{1}{c}{F} &  & Direction & \multicolumn{1}{c}{$\theta_{Fit}$}\\
 & & & [UT] & & \multicolumn{1}{c}{[deg]} & & \multicolumn{1}{c}{km s$^{-1}$} & \multicolumn{1}{c}{[deg]} & & \multicolumn{1}{c}{[deg]} & \multicolumn{1}{c}{[G]} & & & \multicolumn{1}{c}{[deg]}\\
\hline
\endhead

01 & & 1997/01/06 & 15:10 & S18E06 & 15.5 & MC &  136 & 180 & &  25 & 0.8 & & S18W01 & 14.3\\
02 & & 1997/05/12 & 05:30 & N21W08 & 25.2 & MC &  464 & 264 & & 352 & 0.2 & & N01W02 &  4.4\\
03 & & 1997/12/06 & 10:27 & N45W10$^{c}$ & 46.0 & EJ &  397 & 296 & & 315 & 0.5 & & N15W30 & 33.2\\
04 & & 1998/05/01 & 23:40 & S18W05 & 14.8 & EJ &  585 & 126 & & 189 & 0.2 & & S16E14 & 18.3\\
05 & & 1998/05/02 & 14:06 & S15W15 & 18.5 & EJ &  938 & 331 & & 298 & 0.2 & & N08W05 & 13.0\\
07 & & 1998/11/04 & 07:54 & N17W01 & 13.0 & EJ &  523 & 349 & & 324 & 0.1 & & N25W01 & 21.0\\
08 & & 1998/11/09 & 18:18 & N15W05 & 12.6 & EJ &  325 & 338 & & 320$^{d}$ & 1.1$^{d}$ & & N15W05 & 12.6\\
09 & & 1999/04/13 & 03:30 & N16E00 & 21.8 & MC &  291 & 194 & & 301 & 1.9 & & S02W06 &  7.1\\
10 & & 1999/06/24 & 13:31 & N29W13 & 29.7 & EJ &  975 & 335 & & 332 & 0.8 & & N25W15 & 27.2\\
13 & & 1999/09/20 & 06:06 & S20W05 & 27.5 & EJ &  604 &  14 & &  48 &37.0 & & S20W05 & 27.5\\
14 & & 1999/10/18 & 00:06 & S30E15 & 38.3 & EJ &  144 & 184 & &  58 & 9.8 & & S30E15 & 38.3\\
15 & & 2000/01/18 & 17:54 & S19E11 & 17.8 & EJ &  739 &  45 & &  29 & 2.5 & & S10E29 & 29.4\\
16 & & 2000/02/17 & 21:30 & S29E07 & 23.1 & MC &  728 & 184 & & 336 & 0.3 & & S12W02 &  5.5\\
17 & & 2000/07/07 & 10:26 & N04E00 &  0.4 & EJ &  453 & 193 & &  19$^{d}$ & 0.3$^{d}$ & & S17W05 & 21.2\\
18 & & 2000/07/08 & 23:50 & N18W12 & 18.5 & EJ &  483 & 339 & & 279 & 1.7 & & N18W06 & 15.4\\
19 & & 2000/07/14 & 10:54 & N22W07 & 19.0 & MC & 1674 & 273 & & 201 & 5.4 & & N18W14 & 19.5\\
20 & & 2000/07/23 & 05:30 & S13W05 & 18.8 & EJ &  631 & 166 & & 268 & 1.4 & & S13E04 & 18.5\\
21 & & 2000/07/25 & 03:30 & N06W08 &  8.0 & MC &  528 & 168 & & 291 & 8.2 & & S15E04 & 20.6\\
23 & & 2000/08/09 & 16:30 & N20E12 & 18.1 & MC &  702 &  12 & & 159 & 1.0 & & N17E05 & 11.7\\
24 & & 2000/09/16 & 05:18 & N14W07 &  9.8 & MC & 1215 &   3 & & 323$^{d}$ & 6.0$^{d}$ & & N08W07 &  7.0\\
25 & & 2000/10/02 & 03:50 & S09E07 & 17.1 & EJ &  525 & 107 & &  61 &10.0 & & S19E08 & 26.8\\
26 & & 2000/10/09 & 23:50 & N01W14 & 14.9 & MC &  798 & 318 & & 195 & 1.1 & & N20W14 & 19.6\\
27 & & 2000/11/03 & 18:26 & N02W02 &  2.9 & MC &  291 &  57 & & 178 & 0.8 & & N02E05 &  5.4\\
28 & & 2000/11/24 & 05:30 & N20W05 & 19.0 & EJ & 1289 & 313 & & 110$^{d}$ &12.0$^{d}$ & & N30W18 & 33.1\\
29 & & 2001/02/28 & 14:50 & S17W05 & 11.0 & EJ &  313 & 263 & & 342 & 1.9 & & S05W15 & 15.2\\
30 & & 2001/03/19 & 05:26 & S20W00 & 12.9 & EJ &  389 & 184 & & 300 & 0.7 & & N05W10 & 15.6\\
31 & & 2001/04/09 & 15:54 & S21W04 & 15.5 & EJ & 1192 & 211 & & 336 & 2.6 & & S12E01 &  6.1\\
32 & & 2001/04/10 & 05:30 & S23W09 & 19.2 & MC & 2411 & 166 & & 336$^{d}$ & 5.7$^{d}$ & & S23W05 & 17.7\\
33 & & 2001/04/26 & 12:30 & N20W05 & 25.1 & MC & 1006 &  37 & &  62 & 0.5 & & N20W03 & 24.8\\
34 & & 2001/08/09 & 10:30 & N11W14 & 14.7 & EJ &  479 & 255 & & 270 & 1.2 & & N02W18 & 18.5\\
35 & & 2001/10/09 & 11:30 & S28E08 & 35.1 & EJ &  973 & 184 & & 316 & 1.5 & & S28E01 & 34.3\\
36 & & 2002/03/15 & 23:06 & S08W03 &  3.1 & MC &  957 & 309 & & 240 & 0.9 & & N15W01 & 22.2\\
37 & & 2002/04/15 & 03:50 & S15W01 &  9.5 & MC &  720 & 198 & & 335 & 1.1 & & S01W05 &  6.8\\
38 & & 2002/05/08 & 13:50 & S12W07 & 11.1 & EJ &  614 & 229 & & 323 & 2.3 & & S09W09 & 10.6\\
39 & & 2002/05/16 & 00:50 & S23E15 & 25.2 & MC &  600 & 158 & & 360 & 1.3 & & S23E05 & 21.0\\
40 & & 2002/05/17 & 01:27 & S20E14 & 22.3 & EJ &  461 & 145 & &  50 & 0.6$^{e}$ & & S28E20 & 32.0\\
41 & & 2002/05/27 & 13:27 & N22E15 & 27.4 & EJ & 1106 &  35 & & 165 & 4.6 & & N32E20 & 38.2\\
42 & & 2002/07/15 & 21:30 & N19W01 & 14.6 & EJ & 1300 &  45 & & 355 & 4.4 & & N29E15 & 28.6\\
43 & & 2002/07/29 & 12:07 & S10W10 & 18.4 & MC &  222 & 161 & & 203 & 4.4 & & S02W10 & 12.5\\
44 & & 2003/08/14 & 20:06 & S10E02 & 16.7 & MC &  378 &  25 & &  93 & 1.3 & & N12E10 & 11.3\\
45 & & 2003/10/28 & 11:30 & S16E08 & 22.2 & MC & 2495 &  15 & & 160 & 1.1 & & S16E20 & 28.5\\
46 & & 2003/10/29 & 20:54 & S15W02 & 19.7 & MC & 2029 & 190 & & 255 & 4.6 & & S15E05 & 20.2\\
47 & & 2004/01/20 & 00:06 & S13W09 & 12.0 & EJ &  965 & 224 & &  87 & 1.8 & & S25W10 & 22.3\\
48 & & 2004/07/22 & 08:30 & N04E10 & 10.1 & MC &  899 & 210 & & 193 & 1.6 & & N06E05 &  5.1\\
49 & & 2004/11/06 & 02:06 & N09E05 &  7.2 & MC & 1111 &  21 & &  94 & 2.5 & & N07W00 &  3.2\\
50 & & 2004/12/08 & 20:26 & N05W03 &  6.0 & EJ &  611 & 301 & & 233 & 1.2 & & S05W06 &  7.7\\
51 & & 2005/01/15 & 06:30 & N16E04 & 21.0 & EJ & 2049 & 359 & & 113 & 0.6 & & N25W01 & 29.6\\
52 & & 2005/02/13 & 11:06 & S11E09 &  9.9 & EJ &  584 & 129 & & 218 & 5.8 & & S21E19 & 23.6\\
53 & & 2005/05/13 & 17:12 & N12E11 & 18.4 & MC & 1689 &   2 & &  42 & 0.7 & & N05E11 & 13.5\\
54 & & 2005/05/17 & 03:26 & S15W00 & 12.6 & MC &  449 &  54 & & 334 & 2.0 & & N08E05 & 11.5\\
56 & & 2005/07/07 & 17:06 & N09E03 &  6.2 & EJ &  683 &  39 & & 154 & 1.7 & & N12E26 & 27.2\\
57 & & 2005/08/31 & 11:30 & N13W13 & 14.2 & EJ &  825 & 287 & & 191 & 3.4 & & N08W25 & 25.0\\
58 & & 2005/09/13 & 20:00 & S09E10 & 19.0 & EJ & 1866 & 149 & & 100 & 1.3 & & S29E21 & 41.1\\
59 & & 2006/08/16 & 16:30 & S16W08 & 24.0 & EJ &  888 & 161 & & 182 & 0.4 & & S28W01 & 34.7\\
\hline
\multicolumn{15}{l}{$^{a}$\ Data from \citet{gopalswamyetal09} and \citet{mohamedetal12}.}\\
\multicolumn{15}{l}{$^{b}$\ Data from \citet{xieetal12}.}\\
\multicolumn{15}{l}{$^{c}$\ \citet{xieetal12} assumed the source location to be N25W40.}\\
\multicolumn{15}{l}{$^{d}$\ Data value recalculated.}\\
\multicolumn{15}{l}{$^{e}$\ Typo in \citet{mohamedetal12} corrected.}\\
\label{tbl:1}
\end{longtable}
\end{landscape}

%

%
 \begin{acks}
We would like to thank the local organizers of the LWS CDAW meetings in San Diego, USA, and Alcal{\'a} de Henares, Spain. This research was supported by NASA grants NNX10AL50A and NNG11PL10A\@. SOHO is an international cooperation project between ESA and NASA\@.

 \end{acks}

%
%
 \bibliographystyle{spr-mp-sola}
 \bibliography{bibtex_fluxrope}
%
%
%
%

\end{article}
\end{document}